# Leveraging massively parallel reporter assays for evolutionary questions

Irene Gallego Romero[1,2,3,4†*] and Amanda J. Lea[5,6,7†*]

[†]These authors contributed equally
*Corresponding authors' e-mail: amanda.j.lea@vanderbilt.edu and
irene.gallego@unimelb.edu.au

[1]Melbourne Integrative Genomics, University of Melbourne, Royal Parade, 3010, Parkville, Victoria, Australia
[2]School of BioSciences, The University of Melbourne, Royal Parade, 3010, Parkville, Australia
[3]The Centre for Stem Cell Systems, Faculty of Medicine, Dentistry and Health Sciences, The University of Melbourne, 30 Royal Parade Parkville, Victoria 3010, Australia
[4]Center for Genomics, Evolution and Medicine, Institute of Genomics, University of Tartu, Riia 23b, 51010 Tartu, Estonia
[5]Department of Biological Sciences, Vanderbilt University, Nashville, TN 37240, USA
[6]Vanderbilt Genetics Institute, Vanderbilt University, Nashville, TN 37240, USA
[7]Evolutionary Studies Initiative, Vanderbilt University, Nashville, TN 37240, USA

**ABSTRACT**

A long-standing goal of evolutionary biology is to decode how gene regulatory processes contribute to organismal diversity, both within and between species. This question has remained challenging to answer, due both to the difficulties of predicting function from non-coding sequence, and to the technological constraints of laboratory research with non-model taxa. However, a recent methodological development in functional genomics, the massively parallel reporter assay (MPRA), makes it possible to test thousands to millions of sequences for regulatory activity in a single *in vitro* experiment. It does so by combining traditional, single-locus episomal reporter assays (e.g., luciferase reporter assays) with the scalability of high-throughput sequencing. In this perspective, we discuss the execution, advantages, and limitations of MPRAs for research in evolutionary biology. We review recent studies that have made use of this approach to address explicitly evolutionary questions, highlighting study designs that we believe are particularly well-positioned to gain from MPRA approaches. Additionally, we propose solutions for extending these powerful assays to rare taxa and those with limited genomic resources. In doing so, we underscore the broad potential of MPRAs to drive genome-scale functional evolutionary genetics studies in non-traditional model organisms.

**INTRODUCTION**

A major goal in evolutionary biology is to understand why and how adaptively relevant traits differ between individuals and species. Recent advances in genomics have allowed researchers to make rapid progress in this area. In particular, advances in functional genomics have now clarified that changes in gene regulation are important for generating phenotypic variation both within and between species, and frequently contribute to adaptation, speciation, and complex trait evolution[1–6]. Variation in gene regulation also underlies many fundamental biological processes, such as development, tissue differentiation, and the cellular response to environmental stimuli[7–9]. Consequently, there is growing interest in harnessing emerging genomic technologies to address the role of gene regulation in evolutionary processes.

Gene regulatory programs are commonly orchestrated by sequences known as "enhancers". These are *cis*-acting regulatory elements that can be located within, close to, or distal to the genes they regulate (although often within 1 megabase[7,10]), and that can influence gene regulation regardless of their orientation to those genes. Enhancer elements are capable of activating or modulating transcription by recruiting RNA polymerase II and transcription factors, after which they typically contact gene transcription start sites via physical looping[7] (**Figure 1**). Enhancers in the human genome outnumber protein-coding genes by an order of magnitude[7] and allow for the induction of diverse and tissue- or context-specific gene regulatory programs[8]. For example, upon infection, human monocytes upregulate NF-κB/Rel family transcription factors (TFs), which bind enhancers near innate immune genes resulting in mobilization of the cell's defense program[11,12]. Given the context-specific nature of an enhancer's function, mutations in enhancers typically result in less deleterious and less pleiotropic consequences relative to mutations in protein-coding genes, leading some to argue that they may be a preferred substrate of adaptive evolution[13,14]. Indeed, evolutionary turnover of enhancers occurs at much faster rates than other regulatory features (e.g., promoters[15,16]); further, enhancers have been shown to be important for generating morphological novelty in plants and animals[17,18], for maintaining species barriers[5], and for establishing human-specific traits[19–21].

Despite the established significance of enhancers, studying them genome-wide has been difficult, especially outside of humans and model organisms. This is largely because they are difficult to identify from genomic or epigenomic datasets: while enhancers display some predictable sequence features[22,23] and associations with epigenetic marks (e.g., in humans and other vertebrates, they tend to be located in open chromatin regions, hypomethylated, and marked by H3K27ac and/or H3K4me1), these features are not sufficient to predict enhancer activity nor are they exclusive to active enhancers[24,25]. Thus, to confirm the identity, function, and strength of a putative enhancer, experimental validation is required. Such tests commonly involve a "reporter assay", in which a candidate enhancer sequence is cloned into a plasmid containing a minimal promoter and a reporter gene (e.g., GFP, LacZ, or luciferase). The plasmid is then transfected into a cell type of interest, where, if the candidate sequence is indeed an enhancer, it will activate the minimal promoter and result in higher expression of the reporter gene relative to a control construct that only contains the minimal promoter. Such approaches have provided important insight into candidate regulatory elements of evolutionary significance[26–28]. For example, Kvon and colleagues used a reporter assay framework to confirm that snake-specific mutations within the ZRS limb enhancer lead to a reduction in regulatory activity associated with limb loss[26]. While powerful, candidate sequences in this framework are unavoidably tested one-by-one, making the method laborious and impractical when there are many regions of interest, or when discovery of genome-wide patterns is the goal. Recently developed methods, collectively known as "massively parallel reporter assays" (MPRAs), help fill this gap by enabling reporter assay experiments to be carried out in very high-throughput (e.g., testing thousands, hundreds of thousands, or millions of fragments simultaneously; **Figure 1** and **Table 1**). However, due to technical and expertise-related hurdles, MPRAs have thus far been applied mainly to biomedical rather than evolutionary questions. They have also been restricted to a small number of species—namely humans and a few model organisms (i.e., fruit flies[29,30] and mice[31]).

Our goal in this perspective is to showcase how MPRAs can be harnessed to improve our understanding of the generation and evolution of phenotypic diversity across the tree of life. To do so, we first provide an overview of MPRA protocols and their current applications, and in doing so highlight the handful of existing studies that have harnessed MPRA technology for evolutionary questions. We then move to a discussion of study designs that could be leveraged to further address evolutionary questions. We also consider anticipated challenges and potential solutions for expanding MPRA protocols to non-model organisms. We tailor these discussions and recommendations specifically to evolutionary studies, with the aim of highlighting the payoffs of integrating MPRAs into this field.

**OVERVIEW OF MPRA TECHNOLOGIES**

MPRAs grew out of saturation mutagenesis[32,33] and *cis* regulatory element screens[34], which were developed to explore the effects on gene expression of all possible point mutations in a candidate regulatory region. To do so, these protocols linked each of several thousand mutated sequences to a unique barcode. Barcode abundance could be subsequently quantified through RNA-seq, allowing hundreds of different sequences to be tested in a single reporter assay experiment. For example, Patwardhan and colleagues explored the functional impact of every possible mutation in three mouse liver enhancers and found that activity was generally robust to sequence variation: only ~3% of mutations altered enhancer activity by more than two-fold[33]. The protocol innovations that enabled saturation mutagenesis of candidate enhancers (as performed by Patwardhan and colleagues) were quickly applied and optimized to create MPRAs—higher-throughput approaches that could not only test mutagenized sequences of candidate regulatory elements, but also naturally occurring polymorphisms at a genome-wide scale.

MPRAs consist of four main steps. First, DNA sequences of interest are synthesized, each in conjunction with a unique barcode, and are then cloned into a specially engineered plasmid. Second, a basal promoter and a reporter gene are inserted between the sequence of interest and the barcode, such that the barcodes reside in the 3'UTR of the reporter gene. Third, the reporter library is transfected into a cell type of interest, where plasmids containing active enhancers will transcribe the reporter gene and associated barcode. Finally, RNA is extracted from the pool of transfected cells and high-throughput sequencing is used to quantify the barcoded region. In this design, barcode abundance thus scales quantitatively with the regulatory activity of a given tested sequence (**Figure 1A**).

A variation on this design is "self-transcribing active regulatory region sequencing" (STARR-seq), in which the sequence of interest is cloned into the plasmid downstream of a minimal promoter and reporter gene and upstream of a poly-A tail. Consequently, sequences with regulatory activity will interact with the promoter to drive expression of the reporter gene and the sequence itself, such that the abundance of the focal sequence in RNA extracted from cells post-transfection reflects enhancer strength (**Figure 1B**). This approach is similar to the "classic" MPRA design described above and in **Figure 1A**, but circumvents the need for barcodes that tag each tested fragment. Additionally, STARR-seq allows the researcher to use captured, immunoprecipitated, or otherwise selected genomic DNA fragments, as well as randomly fragmented DNA, as input instead of synthesized fragments (**Table 1**).

Many variations on the classic (**Figure 1A**) and STARR-seq flavor (**Figure 1B**) of MPRA designs have been utilized in recent years, with protocol modifications focused on different ways to select DNA input for STARR-seq (e.g., ATAC-STARR-seq[35], ChIP-STARR-seq[36], CapSTARR-seq[37]), integrating MPRA plasmids into the endogenous genome (lentiMPRA[38]), incorporating methyl mark manipulations to test the effects of DNA methylation on enhancer function (mSTARR-seq[39]), or modifying the MPRA framework to study mRNA stability and alternative splicing [40–43]; these changes to the design impact the types of information that can be gained from a given assay (**Table 1**). Additionally, we note that in parallel to the developments we discuss in this review, recent years have seen the establishment of deep mutational scans, which test for effects of all possible mutations in a coding sequence on protein function[44] and are thus analogous to MPRAs in their scale and potential. In some areas of the literature, MPRAs (both the classic and STARR-seq versions) and deep mutational scans have been grouped under the broader header of "multiplexed assays for variant effect" (MAVEs)[45,46]. However, here we focus specifically on assays that consider gene regulation rather than protein function as the output, and we therefore use MPRA rather than MAVE (see **Figure S1** for a terminology hierarchy).

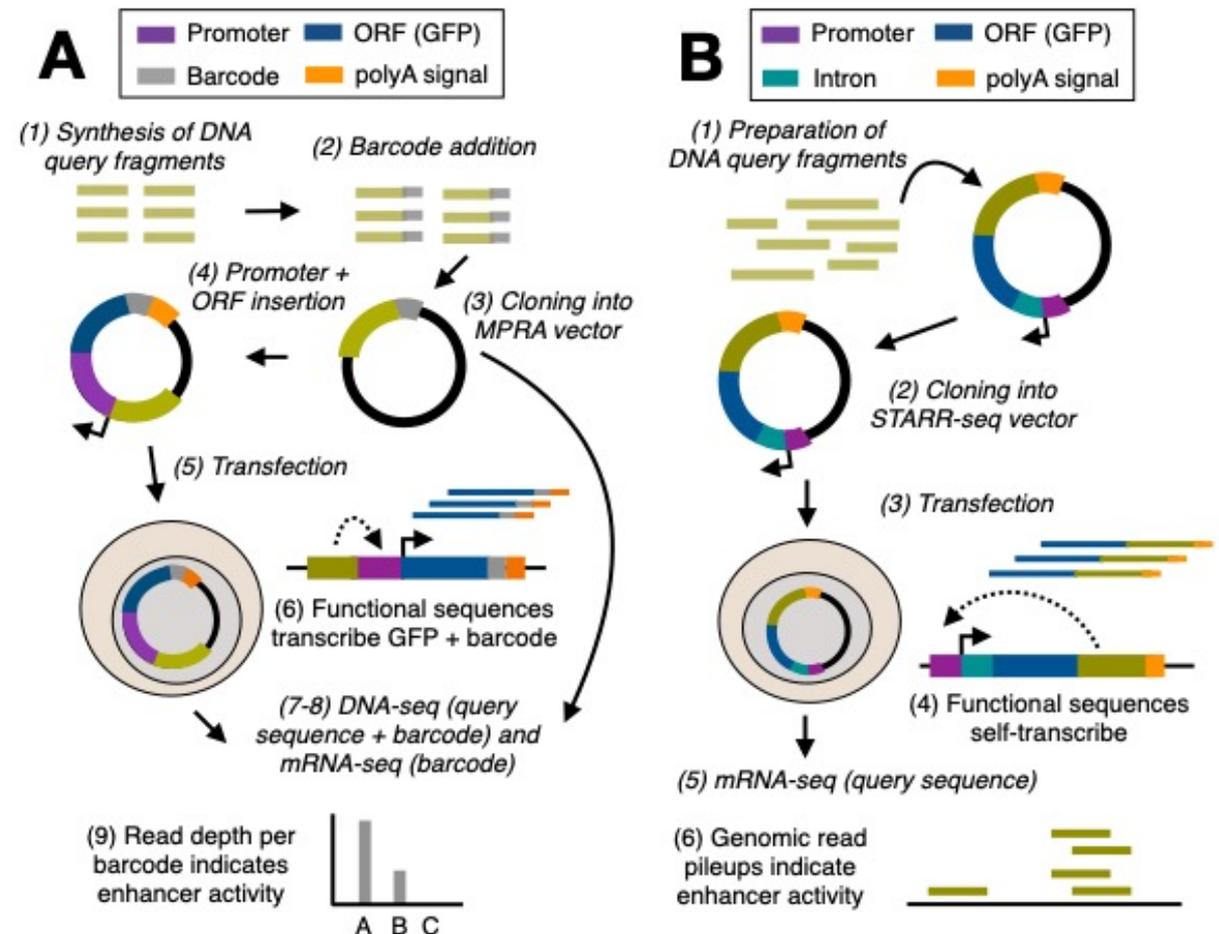

**Figure 1. Overview of MPRA workflows.** (A) (1) In the classic MPRA design, candidate regions of interest are synthesized via large-scale oligosynthesis. (2) The single-stranded DNA is paired with a unique barcode and converted to double-stranded DNA via PCR. (3) The barcoded DNA fragments are then cloned into an empty MPRA reporter vector. Next, the plasmid library is linearized between the barcode and the candidate query sequence, and (4) a minimal promoter and open reading frame are inserted. (5) This plasmid pool is transfected into the desired cell type, where (6) functional enhancer sequences will interact with the promoter to drive transcription of the ORF and the barcode, which is incorporated into each transcript's 3'UTR. Finally, RNA is harvested from the transfected cells and (7) mRNA is isolated and sequenced, along with (8) fragments from the empty MPRA reporter vector step to identify query sequence-barcode associations. (9) Following sequencing, barcode read depths derived from mRNA reflect enhancer activity. (B) (1) In the classic STARR-seq design, sequencing adapters as well as sequences complementary to the STARR-seq vector are added to DNA fragments of interest. (2) This fragment pool is then cloned into the STARR-seq vector upstream of a 3' poly-adenylation signal and downstream of a promoter and synthetic intron (to differentiate spliced mSTARR-seq RNA transcripts from plasmid DNA in downstream PCRs). (3) After transfection into a cell line, (4) inserts that possess enhancer activity interact with the promoter to drive expression of the insert itself. (5) After mRNA purification and sequencing, these inserts can be mapped back to the genome. (6) Read coverage scales with a fragment's ability to drive gene expression.

**CURRENT APPLICATIONS OF MPRAS**

Thus far, studies utilizing MPRAs have been largely focused on biomedical questions addressed in humans and model organisms. In particular, MPRAs have been repeatedly used to tackle a long-standing question in medical genomics: what are the functional pathways linking non-coding regions to disease? MPRAs can shed light on this question by allowing researchers to 1) catalog enhancers, promoters[47,48], and silencers[49] across a variety of disease-relevant human cell types[29,50,51] and cell states[36,52–54] and 2) pinpoint causal alleles within broad disease-associated regions[55,56]. Consequently, MPRAs have been extensively applied to help move beyond the vast catalogs of GWAS-associated loci generated in the past 15 years. For example, Choi and colleagues used an MPRA to characterize the effects of 832 variants in linkage disequilibrium with GWAS hits for melanoma. By pairing MPRA experiments with cis-eQTL mapping and colocalization analyses, the authors were able to identify 4 candidate variants that are likely causal to disease[57]. In another example, Inoue and collegues[58] used a lentiMPRA (**Table 1**) to characterize the dynamics of cis-regulatory element activity across seven timepoints during early neural differentiation. This approach allowed the authors to identify temporally-dependent and independent TFs that regulate neuron development, and to reveal which elements are most active across time, including when cells occupy states of known importance for neurodegenerative disease. Through these studies and many other examples[59–61], MPRAs have proven their utility for uncovering the genetic and mechanistic basis of human disease.

A smaller but growing body of literature has applied MPRAs toward evolutionary questions. For example, MPRAs have been applied to study enhancer evolution in primates[62] and *Drosophila*[30] by comparing the activity of homologous sequences across multiple species. These studies have identified individual regulatory sequences that have gained or lost functional activity across tens of millions of years of evolution, and have also pointed toward generalizable patterns that may characterize such changes. For example, Klein and colleagues linked CpG deamination to significant changes in enhancer activity during primate evolution[62].

MPRAs have also been used to study the function of regions of putative significance to human evolution and human-specific traits. In one instance, Weiss and colleagues explored the effects of ~14k SNPs that are found in modern but not archaic hominins (i.e., Neanderthals and Denisovans)[63]. By functionally assessing both the derived (modern human) and ancestral (archaic hominin) sequence for each region, they were able to show that 23% of regions that had *any* detectable regulatory activity also drove *differential* regulatory activity between modern humans and Neanderthals/Denisovans. These functionally differentiated sequences were enriched near genes involved in traits that also likely differed between modern and archaic humans, such as brain anatomy. Similarly, Uebbing and colleagues[64] as well as Ryu and colleagues[65] both assayed human accelerated regions in neural cell types. Ryu and colleagues coupled MPRA methods with human and chimp induced pluripotent stem cell (iPSC)-derived neural progenitors to compare human, chimpanzee, and intermediate/reconstructed ancestral sequences in equivalent cell types from both species. Using this comprehensive design, they showed that neuronal enhancers with consistent differences in human-chimp activity are almost completely dependent on *cis*-regulatory sequence, with little evidence for interaction with the *trans*-acting cellular environment. Finally, MPRAs have been used to understand the functional consequences of archaic admixture. Jagoda and colleagues[66] as well as Findley and colleagues[67] quantified the regulatory activity of variants introgressed from Neanderthals into the modern human gene pool. Both studies found that many of these variants have causal effects on gene regulation, and thus likely contribute to phenotypic variation today.

**EXPANDING MPRA USAGE IN ECOLOGY AND EVOLUTION**

The examples above highlight the power of MPRAs for improving our understanding of the evolution of phenotypic diversity. While such work so far has been limited to humans and select other taxa, it is highly feasible to apply these approaches to a broader range of species. By applying MPRAs to diverse study designs and organisms, including non-model organisms, many outstanding evolutionary questions could be answered. For instance, in combination with ancestral sequence reconstruction approaches, MPRAs make it possible to test regulatory elements for changes in activity across evolutionary time. In other words, it is possible to assay sequences from both extant and extinct taxa, and thus to explore the evolution of enhancers in general as well as specific enhancer-controlled organismal traits at the gene regulatory level (**Figure 2A**). The strength of this particular approach is unavoidably reliant on the quality and number of existing genome assemblies and is thus not well suited to sparsely sampled phylogenies (we also note there are some caveats in reconstructing ancestral states[68], especially of sequences under selection[69]). However, as the breadth and depth of sequenced genomes increases—for example through large-scale initiatives such as the Vertebrate Genomes Project, Earth Biogenome Project, and DNA Zoo[70,71]—this approach will become more generalizable.

Another possibility is to use MPRAs for fine mapping of functional alleles identified through sequence-based scans for positive selection (**Figure 2B**), analogous to their use to fine-map eQTLs[55] or GWAS hits[56,57]. This could be accomplished by independently testing all SNPs in high linkage disequilibrium within an outlier region or score peak, or, potentially, by tiling across longer elements in small steps to identify functional modules such as key TF binding sites. Such an approach would be extremely useful for addressing a long-standing challenge in evolutionary and population genomics: linking sequence-based measures of adaptation to molecular function and mechanism.

MPRAs could also be applied to understand how genetic interactions (i.e., epistasis and genotype-by-environment interactions) impact phenotypic variation. Notably, genetic interactions have long been thought to be important for complex trait evolution, yet they are notoriously difficult to study because traditional approaches require very large sample sizes to reach statistical robustness[72,73]. MPRAs can be used to make progress in this area. For example, one could assemble a library that includes multiple genotypic versions of a given set of regulatory elements, and then systematically test it 1) within a cell line exposed to different environmental perturbations, 2) within cell lines representing different tissues, or 3) within cell lines derived from the same tissue but from individuals of different genetic backgrounds or species (**Figure 2C**). Doing so would generate quantitative estimates of how varying contexts interact with genetic variation to impact enhancer activity with unprecedented flexibility and resolution.

Importantly, some groundwork has already been laid for these types of study designs. In their study of human-specific variants, for example, Weiss and colleagues tested three different cell types—pluripotent stem cells, osteoblasts, and neural progenitors—and found that most variants were only differentially active between modern and archaic hominins in one of the three cell types[63]. In another example, van Arensbergen and colleagues generated genome-wide MPRA libraries from four individuals included in the 1000 Genomes Project: one person each of Punjab, Japanese, Puerto Rican, and Mende ancestry[74]. They tested ~6 million SNPs in K562 cells (a leukemia cell line) and HepG2 cells (a hepatocarcinoma). Around 30,000 SNPs significantly altered regulatory activity, with ~90% doing so in a cell type-specific manner. Together, these studies point toward a major role for genotype-by-environment effects (in the form of genotype by cell type and state effects) in generating transcriptional variation, at least in humans. We see great potential for expanding this type of work to other species and other types of genetic interactions.

The above examples highlight how MPRAs can be used to catalog the impact of both extinct and extant variation within a population or species at scale. In parallel, deep mutational scans have recently moved beyond a focus on known genetic variation to catalog the effects of all possible mutations within a genomic feature. Taking advantage of error-prone PCR, Kircher and colleauges tested 99.9% of all possible SNPs across 20 different disease-associated regulatory elements to identify those most likely to contribute to their pathogenicity[48]. They found that sequence-based scores of phenotypic impact were generally poor predictors of enhancer activity, pointing to the necessity of functional assays for understanding the consequences of disease-associated variants. To our knowledge these sorts of approaches have not been applied at comparable scale to loci of evolutionary interest, although nothing inherently precludes doing so. Such approaches would be extremely useful for understanding the genotype-phenotype relationship and the landscape of putatively adaptive mutations.

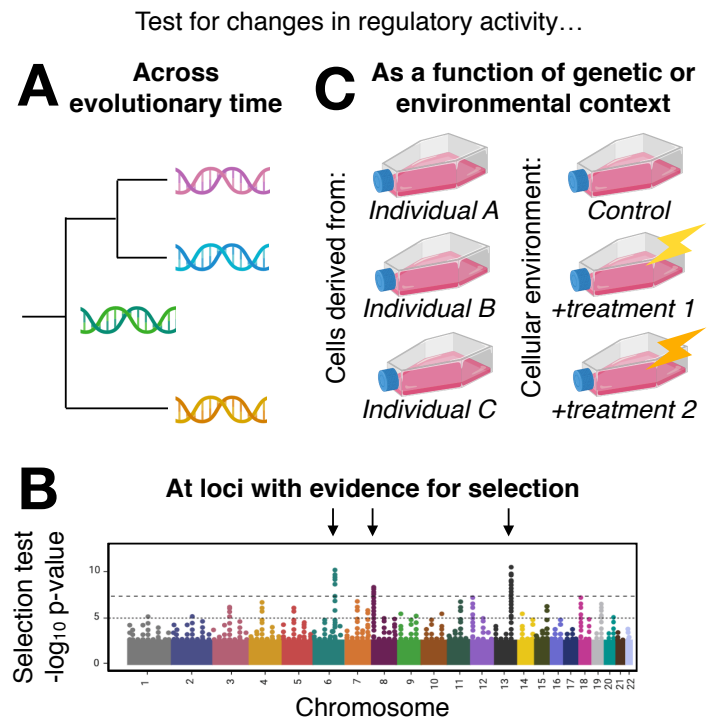

**Figure 2. Study designs for evolutionary questions.** (A) MPRAs can be used to test for changes in enhancer activity across evolutionary time, by assaying orthologous sequences across a phylogeny (pink, blue, and yellow tip lineages) and/or using ancestral sequence reconstruction to assay sequences from extinct taxa (green lineage). (B) MPRAs can be used for fine mapping of functional alleles identified through sequence-based scans for positive selection. (C) MPRAs could be used to understand how genetic interactions, namely epistasis and genotype-by-environment interactions, impact regulatory variation. This could be accomplished by assaying a genetically variable MPRA library across *trans* cellular backgrounds that are either genetically or environmentally diverse.

**CHALLENGES AND RECOMMENDATIONS FOR EXPANDED USAGE**

There are several reasons why MPRA usage has been largely restricted to humans and model organisms thus far. First, we believe there is limited awareness of MPRAs in ecology and evolutionary biology communities, which was a main motivator for writing this review. Second, MPRAs are complex assays and require access to specialized equipment and know-how to carry out. However, most of the equipment (e.g., biosafety cabinets, incubators, electroporators) is extremely common in molecular- or genetics-focused departments and likely already exists at most institutions. Further, several detailed MPRA protocols are now publicly available[75,76] (**Table S1**), making it increasingly feasible for researchers with diverse expertise to apply these assays. Third, in addition to specialized equipment and know-how, MPRAs also require 1) a high-quality genome sequence and/or large amounts of genetic material, depending on the study design, and 2) a relevant primary cell pool or immortalized cell line for transfection. These technical constraints have likely hindered the widespread adoption of MPRAs, although we believe both can be overcome.

One of the main challenges associated with expanding usage of the classic MPRA design is that it relies on large-scale oligosynthesis of known genomic sequences. Thus, a reference genome is required. Reference genomes are increasingly available for most study organisms, as well as increasingly feasible to generate *de novo*[77,78]. Alternatively, a subset of the genome could be sequenced at much lower cost using methods like RAD-seq[79,80], as well as methods that specifically target gene regulatory elements (e.g., ChIP-seq[81,82] or ATAC-seq[83]), which can then be used to refine the list of testable sequences. A more general challenge for MPRAs is that oligosynthesis is limited in both capacity and sequence length: commercial providers rarely synthesize fragments longer than 300 bp. This means that most classic MPRAs test short sequences, or require sliding window designs to examine larger ones, introducing additional complexity during analysis. While 300 bp are enough to capture, for example, specific TF binding sites and local interactions, many complete enhancer elements (or sets of nearby enhancer elements) are larger than 300 bp. Indeed, studies thus far demonstrate increased power to detect enhancer activity when query fragments are larger, as well as a general impact of fragment length on downstream assay output[39,84].

An alternative approach is to use STARR-seq family methods (**Table 1**) to support testing of larger fragments. Such approaches can leverage either sequence-capture or other methods to target DNA fragments of interest or random shearing to cover an entire genome. Either design requires access to large amounts of starting genetic material (e.g., a few[29] to hundreds[52] of micrograms of DNA, or potentially reliance on whole genome amplifications[85]); this input requirement may pose challenges when working with rare samples or endangered species. However, once a plasmid library is generated, it can be easily renewed via bacterial transformation with minimal loss of diversity[39]. Therefore, while it may be challenging to collect micrograms of DNA for some species, for many study designs this obstacle only needs to be overcome once; the resulting plasmid library can then support multiple experiments and even be shared across the scientific community. Depending on the questions, it may also be worthwhile to pool smaller amounts of material from many individuals to create a single library of genetically diverse regulatory elements[86].

Once a plasmid library is assembled, an unavoidable challenge for many studies will be the need for a cell line that can be grown at scale, efficiently transfected, and is representative of the species and tissue of interest. The first two requirements are intimately linked to the number of sequences that can be tested in a given assay. This is because each sequence of interest must be assayed independently multiple times to achieve robust statistical power. Recent recommendations in the field for classic MPRA designs are to ensure that every sequence is represented by 50-100 independent barcodes, with multiple observations of each barcode[55]. With these numbers, testing just 20,000 sequences may require successful transfection of 10-20 million cells, with larger starting cell amounts needed since transfection efficiency is never 100%. For STARR-seq designs, recommendations are to successfully transfect ~60 or ~300 million cells for focused versus genome-wide screens, respectively[76].

These cell numbers can be prohibitive in the case of hard-to-transfect, terminally differentiated, or non-proliferative cell types, or when working with rare samples or non-model species. Indeed, commercially available cell lines with pre-optimized growth and transfection protocols are for the most part limited to humans and model organisms, though a growing number of commercially available products are available for other species (**Figure 3**). In some cases, it may be feasible to use modified MPRA protocols appropriate for hard to transfect cell types and/or limited cell quantities[65], or to derive new cell lines for non-model species[87]. In other cases, a better solution may be to use a cell line from a closely related species as a proxy (e.g.,[30,62]). This design assumes a conserved *trans* environment since the split of the focal and cell line species, but there is strong evidence that TF expression, structure, and specificity to binding motifs are well-conserved across long evolutionary time scales[88–90]. For instance, we reanalyzed gene expression data from human, gorilla, chimpanzee, orangutan, and macaque lymphoblastoid cell lines[91] (LCLs) and compared TF expression levels between humans and each of the other species. We found that TF expression levels in LCLs are highly conserved across species pairs spanning ~6 to ~26 million years of evolutionary divergence ($R^2$ for pairwise comparisons=0.66-0.76; **Figure S2**). It is also worth highlighting that one MPRA study so far, in humans and chimpanzees, has already shown that the overwhelming majority of human-chimpanzee species differences in enhancer activity arise from the query fragment sequence itself rather than the species-specific cellular environment; in this study, *trans* effects generated differences in activity for <1% of regulatory elements[65]. Thus, several lines of evidence suggest that the easiest solution for non-model organism researchers is to use an existing cell line from a closely related species, and that this choice will have minimal effects on evolutionary inferences.

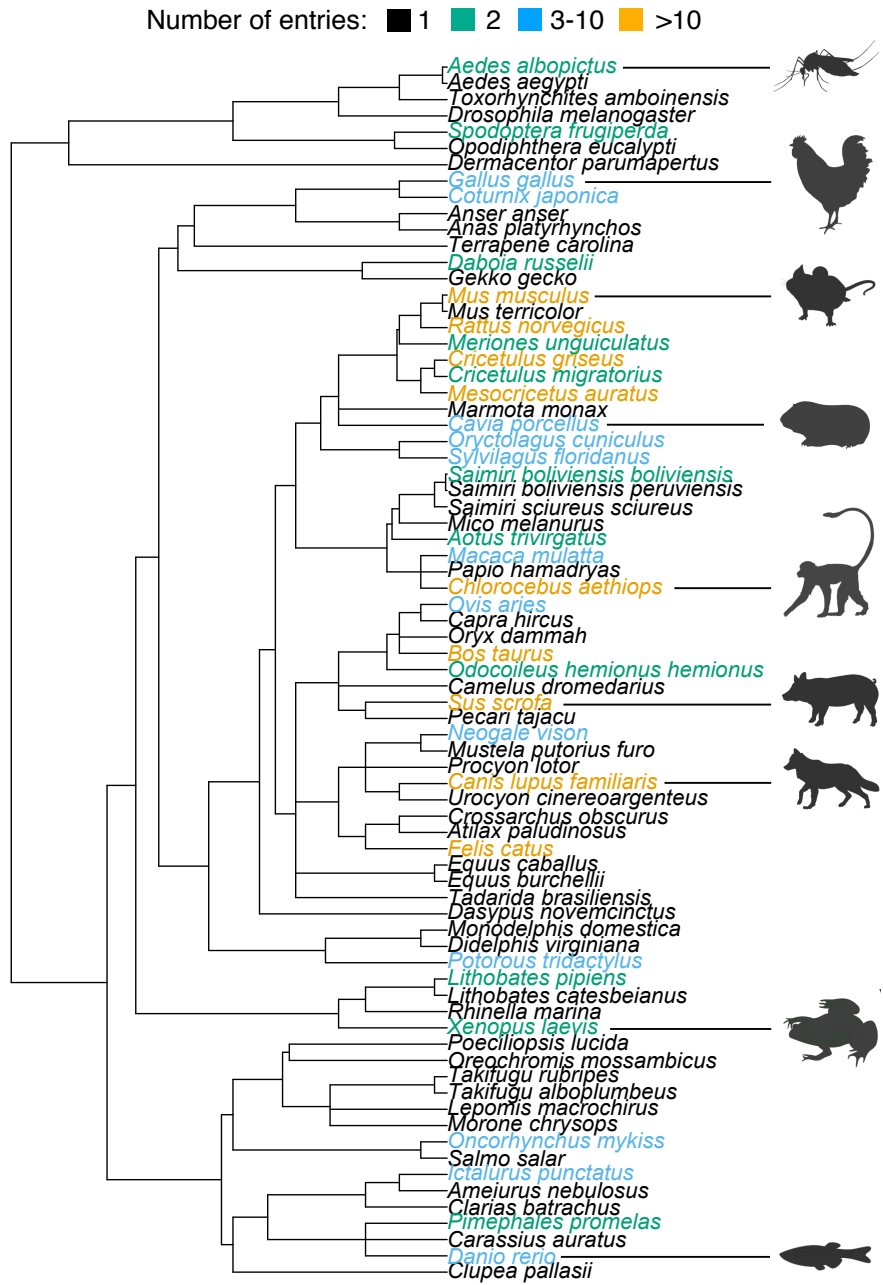

**Figure 3. Animal species currently represented in the American Type Culture Collection (ATCC) catalog.** Phylogenetic tree of all species with entries in the ATCC catalog, color coded by the number of entries. Select species are represented with cartoons to orient the reader to where broad taxonomic groups fall on the phylogeny.

## FUTURE DIRECTIONS AND CONCLUSIONS

Moving forward, there are two areas where emerging research from non-evolutionary fields will soon benefit evolutionary biologists, and in turn catalyze research in this area. First, there is a growing awareness of the potential MPRAs hold, and a growing community drive to develop standards to facilitate community adoption and data reuse. For instance, MaveDB provides a resource for deposition of results from MPRAs (and other types of MAVEs) under a standardized format[92]. Similarly, the nascent Alliance of Variant Effects (AVE) seeks to build an atlas of all possible variants in disease-related functional elements in the human genome[93]. These existing data collections could be mined for inferences about human evolution, but more broadly these efforts signal that public, standardized databases will be the norm going forward, and will surely benefit the evolutionary community as they are expanded to a wider range of species. Second, MPRAs have recently motivated bioinformatic and statistical tool development[94–98], which could aid non-model organism researchers as more MPRA data are generated for these species. For example, MPRA data can be coupled with machine learning approaches[99–108] to predict gene expression and regulatory structure from genomic sequence alone. For instance, MPRA-DragoNN[99] uses densley tiled MPRA data[51] to predict DNA sequence features that impact expression differences, while DeepSTARR[100] uses STARR-seq data to predict enhancer location and function genome-wide. These tools could allow non-model organism researchers to bioinformatically generate genome-wide regulatory maps from a focused MPRA dataset, or potentially, from one generated for a closely related species.

Like most other genomic technologies, MPRAs were first optimized in systems with extensive genomic resources (i.e., humans and model organisms). However, for evolutionary biologists, these approaches often become most exciting once they are expanded to a more diverse set of species and contexts—even if these extensions come with caveats and challenges. We believe that the biological insights to be gained from applying MPRAs to diverse organisms, environments, and study designs have substantial potential for addressing evolutionary questions. In particular, we believe MPRAs will soon expand our ability to interpret and annotate the genomes of non-model organisms, as well as our understanding of how gene regulation contributes to adaptive evolution and phenotypic diversity. The already demonstrated significance of MPRAs in the biomedical sciences suggests that, in the coming years, we can expect an equivalent wealth of insights drawn across a broad range of taxa and evolutionary questions.

**ACKNOWLEDGEMENTS**

We thank members of the Gallego Romero and Lea labs, as well as Jenny Tung and Christopher Vockley, for extremely valuable feedback. We thank Jenny Tung for the bones of Figure 1. A.J.L. is supported by a Dean's Faculty Fellowship Award from Vanderbilt University.

**AUTHOR CONTRIBUTIONS**

I.G.R. and A.J.L. conceived the study and wrote the manuscript.

**COMPETING INTERESTS**

The authors declare no competing financial interests.

**TABLES**

**Table 1. An overview of different MPRA approaches.**

| Assay | Summary |
|---|---|
| *"Classic" methods* | |
| MPRA[25,33,55,109] | DNA sequences of interest are each synthesized in conjunction with a unique barcode and cloned into a plasmid upstream of a promoter, reporter gene, the unique barcode, and a poly-A tail. Sequences with regulatory activity drive expression of transcripts that include the barcode, such that barcode abundance in RNA extracted from transfected cells reflects enhancer strength. |
| STARR-seq[29] | Sequences of interest are cloned into a plasmid downstream of a minimal promoter and reporter gene and upstream of a poly-A tail. Sequences with regulatory activity drive expression of transcripts that include the sequence itself, such that the abundance of the focal sequence in RNA extracted from transfected cells reflects enhancer strength. |
| *Elaborations on the classic MPRA design* | |
| lenti-MPRA[38] | Lentivirus is used to integrate MPRA libraries into the genome, thereby circumventing concerns that episomal reporter assays carried out via transient transfection may not reflect gene regulatory processes that take place in a native chromatin context. The cell-type range of lentivirus transduction is also much broader than transient transfection, opening the door to experiments in hard to transfect cell types. |
| AAV MPRA[110] | MPRA libraries are packaged into an adeno-associated virus (AAV) for transfection. AAV is a nonpathogenic virus commonly used for gene therapy studies, and permits transfection into a wide range of tissues, including post-mitotic tissues and tissues that are hard to transfect with traditional chemical or electrical methods. Unlike DNA delivered by lentivirus, the AAV-delivered DNA remains almost exclusively episomal. |
| saturation mutagenesis-based MPRA[48] | To test the functional effects of thousands of mutations in a candidate regulatory element, error-prone PCR is used to introduce sequence variation and to incorporate random sequence tags. These constructs are then assayed via the MPRA design to pinpoint SNPs that affect regulatory activity. |
| *Elaborations on the classic STARR-seq design* | |
| STAP-seq[111] | Rather than measuring the activity of many candidate enhancers in the presence of a given minimal promoter, STAP-seq measures the responsiveness of many candidate promoters in the presence of a given enhancer. Promoter candidates are cloned downstream |

|  | of a strong enhancer and upstream of an ORF and poly-A tail. If a candidate fragment is capable of initiating transcription, it will produce reporter transcripts that start with the promoter candidate sequence wherever the TSS was initiated. |
|---|---|
| UMI-STARR-seq[76] | This protocol introduces unique molecular identifiers (UMI) prior to post-transfection amplification of cell-extracted mRNA. The UMIs allow the researcher to account for PCR duplicates in downstream analyses, and are recommended especially for low complexity input libraries. |
| ChIP-STARR-seq[36] | Open chromatin regions are incorporated into a DNA library, which is then assayed via STARR-seq. |
| PopSTARR-seq[86] | Regions of interest are amplified from DNA derived from many unique individuals. These genetically diverse products are then pooled and used as the input for STARR-seq. |
| ATAC-STARR-seq[35] | Open chromatin regions are incorporated into a DNA library via ATAC-seq[83], and these elements are then assayed via STARR-seq. This design allows the researcher to preferentially test the activity of putative regulatory elements found within open chromatin in a given cell type. |
| BiT-STARR-seq[112] | Oligos covering each of the alleles for a set of SNPs of interest are synthesized and incorporated into STARR-seq experiments to test for allele-specific expression. UMIs are also added during cDNA synthesis to account for PCR duplicates. |
| mSTARR-seq[39] | STARR-seq style plasmid pools are constructed using a CpG free reporter vector that retains the same functionality. Enzyme treatment is then used to create methylated and unmethylated versions of the plasmid pool, which can be assayed to identify regulatory sequences as well as methylation-dependent regulatory sequences. |
| CapSTARR-seq[37] | Putative enhancers are selected from genomic DNA using hybridization capture-based target enrichment. Captured regions are then assayed via STARR-seq, allowing the researcher to test a targeted set of fragments without relying on oligo synthesis. |

Supplementary Materials for Gallego Romero and Lea, "**Leveraging massively parallel reporter assays for evolutionary questions**"

This PDF contains:

Supplementary Methods
Supplementary Figure 1
Supplementary Figure 2
Supplementary Table 1
Supplementary References

**Supplementary Methods**

To understand the existing cellular resources for non-human species, we downloaded the ATCC catalog for all animal species excluding humans. We then used the R package taxize[1] to download taxonomic hierarchical information for each species using the option for querying the NCBI database. In a few cases, we corrected spelling errors or out of date genus names to recover taxonomic information.

To understand how conserved TF expression is through evolutionary time, we reanalyzed gene expression data from human, gorilla, chimpanzee, orangutan, and macaque lymphoblastoid cell lines[2] and compared TF expression levels between humans and each of the other species. To do so, we used the R package biomaRt[3] to identify orthologous protein coding genes for each human and non-human primate species pair. We calculated the species-specific mean $\log_2$ TPM value and removed genes that were not expressed in either species. Finally, we compared gene expression levels across all protein coding genes as well as for TF genes alone using Pearson's $R^2$. Results are plotted in Figure S1.

All analyses were conducted in R version 4.1.2. Figures 2 and 3 use images from BioRender.com.

**Supplementary Figures**

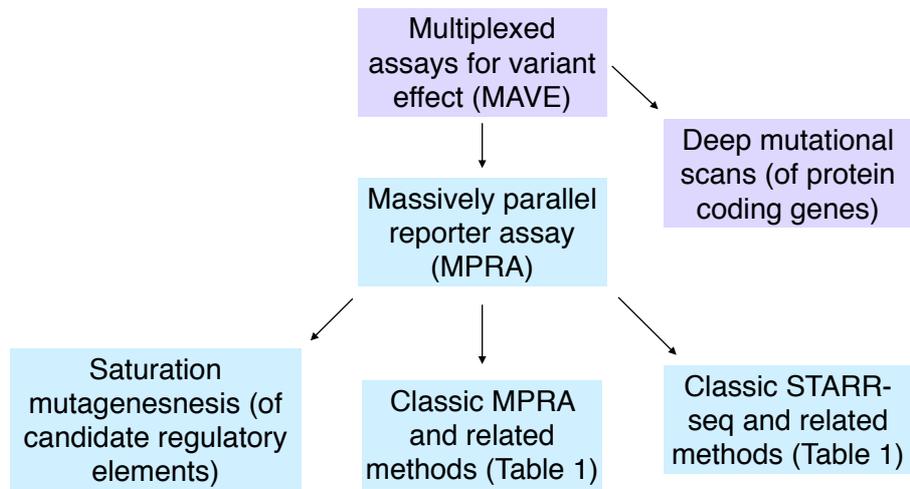

**Supplemetary Figure 1. Hierarchy of terminology used in this review**. Terms in blue are the focus of this review.

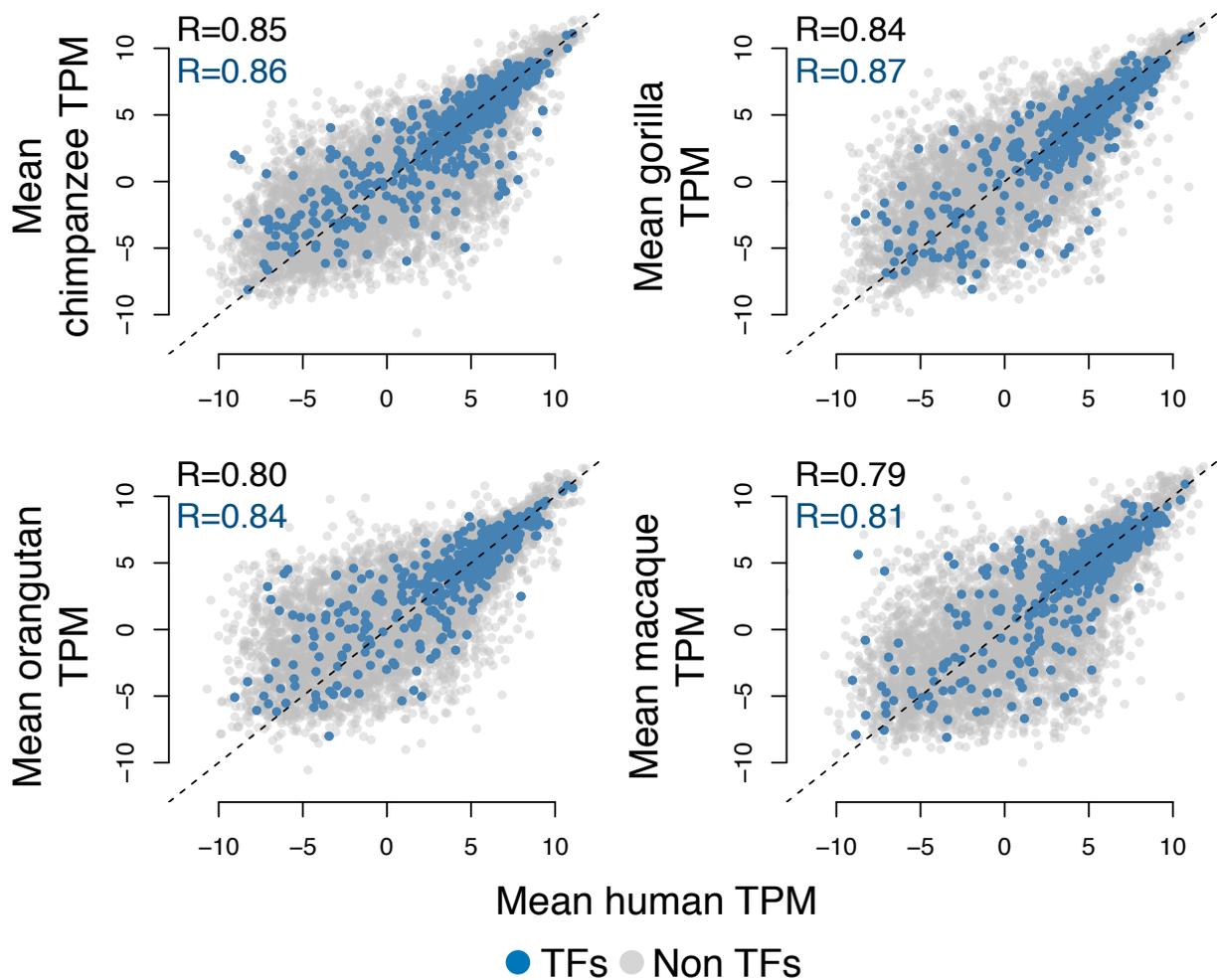

**Supplementary Figure 2. Conservation of transcription factor (TF) expression across primate species**. x and y-axes represent the mean $\log_2$ gene expression values for a given species (in terms of transcripts per million, TPM). Only orthologous genes for a given species pair are included, and each dot represents a gene (colored according to whether it is or is not a TF in the TRRUST database of mammalian transcription factors[4]). Dotted lines represent x=y.

**Supplementary Table 1. Available step by step protocols.**

| Assay | Protocol link |
|---|---|
| ATAC-STARR-seq | https://www.protocols.io/view/atac-starr-seq-5jyl89rorv2w/v1 |
| STARR-seq and UMI-STARR-seq | https://currentprotocols.onlinelibrary.wiley.com/doi/full/10.1002/cpmb.105 |
| mSTARR-seq | http://www.tung-lab.org/protocols-and-software.html |
| CapSTARR-seq | https://protocolexchange.researchsquare.com/article/nprot-4333/v1 |
| MPRA | https://www.protocols.io/view/massive-parallel-reporter-assay-mpra-kxygxpmkwl8j/v1 |
| MPRA | http://noonan.ycga.yale.edu/noonan_public/Uebbing_Gockley_MPRA/Extended_Methods.pdf |

**Supplementary References**